# Stabilization of Magnetic Skyrmions on Arrays of Self-Assembled Hexagonal Nanodomes for Magnetic Recording Applications

Felipe Tejo,[1,4,5] Denilson Toneto,[2] Simón Oyarzún,[1,4] José Hermosilla,[1] Caroline S. Danna,[1] Juan L. Palma,[3,4] Ricardo B. da Silva,[2] Lucio S. Dorneles,[2] and Juliano C. Denardin[1,4,*]

[1]*Departamento de Física, Universidad de Santiago de Chile, Santiago, Chile.*

[2]*Departamento de Física, Universidade Federal de Santa Maria, UFSM, Santa Maria, RS, Brazil.*

[3]*Escuela de Ingeniería, Universidad Central de Chile, Santiago, Chile*

[4] *CEDENNA, Universidad de Santiago de Chile, Santiago, Chile.*

[5] *Instituto de Ciencia de Materiales de Madrid, CSIC, Cantoblanco, 28049 Madrid, Spain.*

**ABSTRACT:** Magnetic skyrmions are nontrivial spin textures which resist external perturbations, being promising candidates for the next generation recording devices. Nevertheless, a major challenge in realizing skyrmion-based devices is the stabilization of ordered arrays of these spin textures under ambient conditions and zero applied field. Here, we demonstrate for the first time the formation and stabilization of magnetic skyrmions on arrays of self-assembled hexagonal nanodomes taking advantage of the intrinsic properties of its curved geometry. Magnetic force microscopy images from the arrays of 100 nm nanodomes showed stable skyrmions at zero field that are arranged following the topography of the nanostructure. Micromagnetic simulations are compared to the experiments to determine the correlation of the domain textures with the topography of the samples. We propose a simple method to nucleate and annihilate skyrmions, opening the possibility for ultra-dense data storage based on the high stability and low energy consumption of the skyrmionic textures.

**INTRODUCTION**

The confinement of stable magnetic textures in nanostructures is fundamental to modern magnetic recording technology[1]. Additionally, there is a special class of magnetic materials, whose crystallographic lattice lacks inversion symmetry, in which it is possible to stabilize chiral spin configurations such as magnetic skyrmions[2,3]. These magnetic states are generally explained by the existence of a type of interaction known as Dzyaloshinskii-Moriya interaction[4-7] (DMI), and have interesting topological properties that contribute to their stability and to energy efficiency in controlling their movement[8], as they can be displaced by means of relatively low spin-polarized currents[8,9]. Magnetic multilayers (ML) with ferromagnetic/heavy metal (FM/HM) interfaces and broken structural inversion symmetry offer the possibility of tuning the magnetic easy axis from in-plane to perpendicular-to-plane, what makes them a rich playground to investigate skyrmionic textures[10-13]. Recent results show skyrmions being stabilized on MLs with perpendicular magnetic anisotropy (PMA)[4] and even on films with in-plane magnetic anisotropy[13]. However, a major challenge in realizing skyrmion-based devices is their stabilization under ambient conditions and at zero applied field[14].

Geometrical confinement is important for skyrmion nucleation and stabilization, and some strategies have been tested to obtain confined arrays of skyrmions by the fabrication of patterned nanostructures by lithography[15] or by focused ion beam[16], both involving high-cost and complicated processing. A different approach to obtain nanopatterned films is to use nanoporous alumina membranes (NAMs)[17]. Recent reports have shown that films deposited on top of arrays of ordered nanostructured membranes with nanopores[18] or nanodomes[19] can retain PMA and present interesting magnetic and magnetotransport properties[20], some of which arising from their geometry[21].

One of the most interesting properties exhibited by low-dimensional curved systems is the emergence of a curvature-induced anisotropy and an effective Dzyaloshinskii–Moriya interaction (DMI)[22-24], leading to magnetochiral effects[25-27] and topologically induced magnetization patterning[24,28]. Recently, Carvalho-Santos et al. reported that the presence of curvature increases skyrmion stability[29]. Consequently, well-controlled nanodome synthesis emerges as an alternative to control specific magnetic states by adjusting geometrical characteristics. Although magnetic nanodomes can be produced by the deposition of multilayers onto the barrier layer of NAMs, there are no reports of their use to stabilize and control magnetic skyrmions.

In this work we report on the production of an ordered array of Pt/Co/Ta multilayers deposited onto nanodomes with 100 nm of diameter, and we present a direct observation of the nucleation, stability and annihilation of magnetic skyrmions at low applied fields in positions correlated to the structural arrangement of the nanodomes. Micromagnetic simulations were compared to the experiments and determine the correlation between the position of the domain textures and the topography of the samples. We propose a skyrmion-based data storage device that incorporates the nucleation and annihilation of individual skyrmions, as well as the control of their position by application of magnetic field.

**EXPERIMENTAL SECTION**

The procedure to fabricate the samples on nanodomes is a three-stage method where the film is deposited onto the barrier layer of NAMs[20]. The first step consists of the fabrication of NAMs by double anodization technique[17] and subsequent removal of the remaining aluminum (Al) by chemical etching, exposing the bottom oxide barrier layer, as shown in the sketch of Fig.1(a). The resulting pore diameter was controlled by the type of acid and the applied voltage during anodization[20]. The inter-pore distance is related to the nanodome diameter in the oxide barrier layer, and our sample set is composed of hexagonal arrangements with 100 nm diameter, as can be observed in Fig. 1(b). The membranes with nanodomes have been subsequently used as substrates for the deposition of the Pt/Co/Ta multilayers. The multilayer stacks Ta (4.7 nm)/[Pt(4 nm)/Co(1.3 nm)/Ta(1.9 nm)]$_{x15}$ were deposited by DC magnetron sputtering at room temperature. The base pressure in the chamber was $8\times10^{-7}$ Torr, and the Ar pressure during deposition was kept at 3 mTorr using a 20 sccm Ar flow. Reference multilayers were also deposited onto a flat Si substrate (reference ML). The static magnetic properties of the multilayers have been measured at room temperature in a 5T-VSM from Cryogenic Ltd. Magnetic domain pattern images of the multilayers were acquired by an ezAFM from Nanomagnetic Instruments operating in the dynamic MFM mode. We used Multi75-G (75 kHz) tips from Budget Sensors, which are coated by a cobalt alloy presenting magnetic moment and coercivity of roughly $10^{-16}$ Am$^{-2}$ and 0.03 T, respectively. The images were acquired at room temperature with a tip-surface distance of about 60 nm. Micromagnetic simulations have been performed with the Mumax3 GPU-accelerated program[30] to solve the time-dependent Landau-Lifshitz-Gilbert (LLG) equation. In the simulation, a $1 \times 1$ µm$^2$ square system, applying the 2D periodic boundary conditions and a mesh size of $4 \times 4 \times 4$ nm$^2$ with 5 repetitions in *z*, was used to simulate the multilayers. The arrays of nanodomes have been built from hollow half spheres with external diameters of 100 nm and repeated in hexagonal arrays.

**RESULTS AND DISCUSSIONS**

**Characterization of the samples**. Fig. 1(b) shows the SEM image from the film deposited onto the barrier layer of NAMs with 100 nm of inter-pore distance, where it is possible to observe the film homogeneity and the regular hexagonal arrangement of the domes. Fig. 1(c) and Fig. 1(d) show the normalized hysteresis loops from the continuous flat film (reference ML) and from the film on nanodomes, respectively. The hysteresis curves from the reference ML present an out-of-plane anisotropy. The bow-tie shape of the out-of-plane hysteresis loop suggests a strong dipolar interaction that, with the DMI, may help to stabilize skyrmions[10,11]. The out-of-plane hysteresis curve from the film on nanodomes does not have the same aspect. The bow-tie shape disappears, and the saturation field is comparable to the one observed in the in-plane hysteresis curve, indicating that out-of-plane and in-plane anisotropies are comparable. The out-of-plane hysteresis curve from the film on nanodomes also shows an enhancement of coercivity and remanence, when comparing to the curve from the reference ML. This can be considered an evidence that the domain wall propagation is affected by the presence of several pinning centers generated by the topological characteristic of the nanodomes[19,20]. As the curved topography of the nanodomes also affects the magnetic anisotropy, both the anisotropy field $H_k$ and the anisotropy constant ($K_{eff} = M_s H_k/2$) values decrease for the film on nanodomes; the anisotropy constant changes from $K_{eff} = 0.21$ MJ/m$^3$ to $K_{eff} = 0.17$ MJ/m$^3$. Recent works have shown that the magnetic anisotropy plays an important role in determining the topological spin configuration[10], as samples with competing anisotropies showed higher skyrmion densities[11]. Analytical calculations of dome structures also predicted a decrease in effective anisotropy with increasing curvature and concluded that the presence of curvature increases skyrmion stability[29].

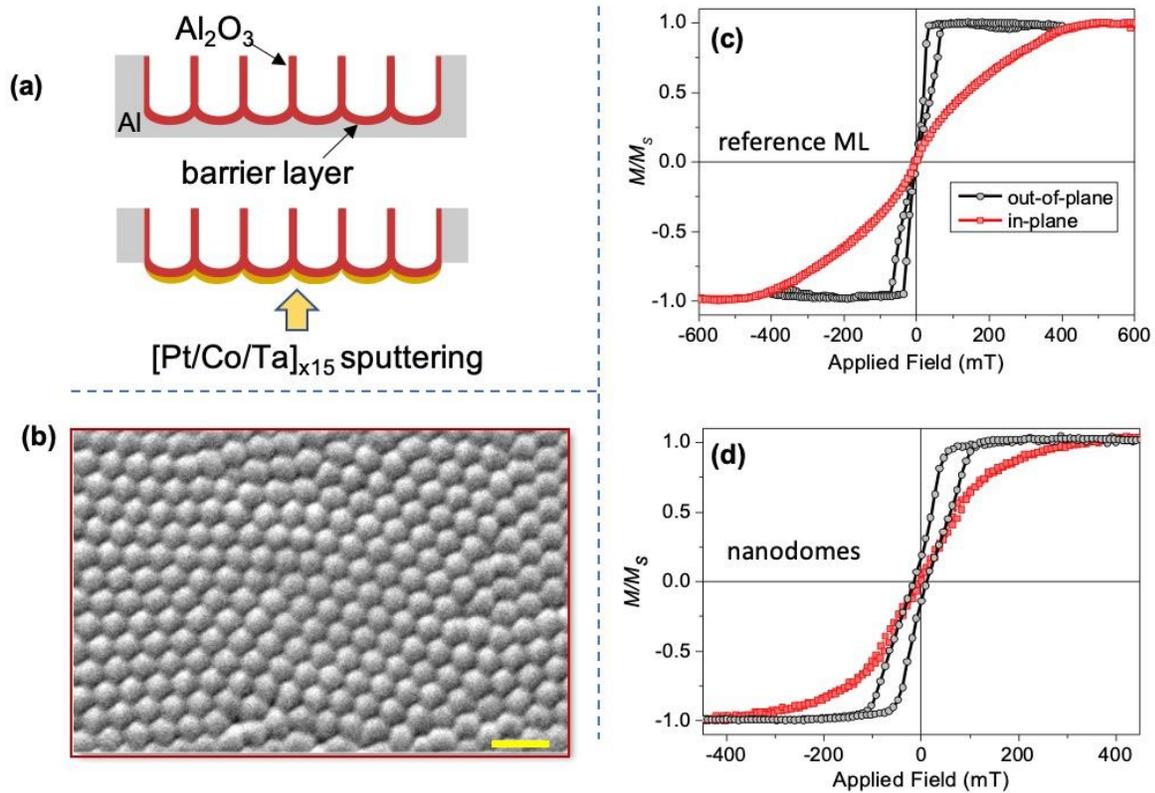

Fig. 1: (a) Schematic representation of the fabrication of samples. After anodization and removal of Al, multilayers are deposited by sputtering in the bottom of the membrane. (b) SEM image of the nanodome array with 100 nm (scale bar is 200 nm). (c-d) Out-of-plane and in-plane hysteresis curves of the Pt/Co/Ta films deposited on (c) reference multilayer and (d) on nanodomes.

We used magnetic force microscopy (MFM) images to observe the magnetic configurations of the demagnetized state of the samples. The MFM images from the reference ML and from the film on nanodomes are shown in Fig. 2 (a) and Fig. 2(c), respectively. As expected for films with both PMA and DMI[12], a typical maze domain structure is clearly seen in Fig.2(a), in a demagnetized state obtained by application of an alternating magnetic field with decreasing intensity. The average domain width was determined using line scans on different parts of the MFM image, and a representative profile is shown in Fig. 2(b); the domain width measured at half maximum of each peak is $d_w \approx 250$ nm. In the MFM image of Fig. 2(c) from the film on nanodomes, it is clearly seen that the topography of the nanodomes strongly affects the domain configuration, as the stripe domains become smallest and thinnest. In Fig. 2(d) it is shown in detail the presence of a mixture of thin stripe domains, small circular domains, and ordered chains of circular domains following the topography of the nanodomes; as shown by the MFM profile in Fig. 2(e), with an average periodicity of 100 nm. These results indicate that the curved substrate effectively changes the domain configuration in the ML and show that the width of the stripe domains can be reduced by the nanomodulation of the substrate.

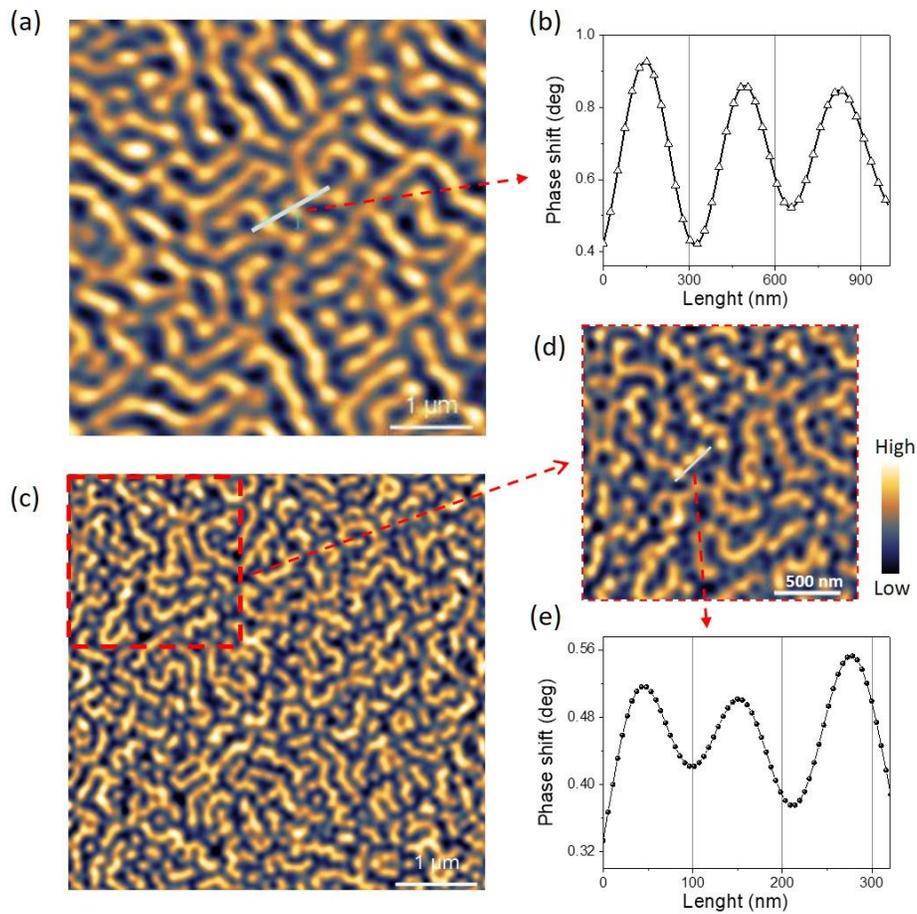

Fig. 2. (a) MFM image of the reference ML in the demagnetized state. (b) Magnetic profile taken on the MFM image (light gray line) used to obtain the width of the domains. (c) MFM image of the nanodomes obtained in the demagnetized state and (d) a magnified detail from the square red region where the magnetic profile of the aligned chain of circular domains, marked by a gray line, is shown in (e).

A feature of the nanodome systems is their geometrically broken symmetry which leads to interesting physical properties, and important for the study of magnetic configurations[19,20]. It is well known that in curvilinear systems there appear two effective magnetic interactions: curvature-induced effective anisotropy and curvature-induced effective Dzyaloshinskii-like interaction[22]. The emergence of these interactions gives rise to magnetochiral effects that allow the stabilization of non-trivial magnetic textures (e.g. skyrmions or chiral domain walls). An example is the exchange interaction, which contains three components of different symmetries in curvilinear coordinates: $\mathcal{E}_{ex} = \mathcal{E}_{ex}^0 + \mathcal{E}_{ex}^A + \mathcal{E}_{ex}^D$, where $\mathcal{E}_{ex}^0$ is a regular isotropic part of exchange interaction, which has the form similar to the one in a planar film, $\mathcal{E}_{ex}^A$ describes a curvature-induced biaxial anisotropy and $\mathcal{E}_{ex}^D$ is a curvature-induced extrinsic DMI. Similarly, it is possible to restructure all the magnetic energy terms containing spatial derivatives (for a detailed review of mathematical treatment, please see reference[31]). The extended hexagonal lattice produces a non-local effect on the demagnetized state[19,20]. Additionally, the curvature effect produces a re-scaling of the energy due to the additional terms that appear (anisotropy and DMI), which produces thinner stripe domains.

**Skyrmions nucleation and stability on nanodomes.** The nucleation of skyrmions on magnetic multilayers is a somewhat tricky combination of specific magnetic parameters and the application of magnetic field sequences[32]. Recent results have shown that the labyrinth domains can be ordered in stripe domains with in-plane fields, and then the stripe domains can be broken into skyrmion arrays when the field is applied at some angle relative to the sample plane[12]. Isolated skyrmions have been also nucleated when strong out-of-plane magnetic fields have been applied[10,11], and stabilized at remanence after out-of-plane field sequences (that produce a specific irreversible magnetization process) are applied[32]. In a recent work, we used first order reversal curves (FORC) Hall analysis to determine the necessary magnetic field to be applied in order to nucleate and stabilize skyrmions in Pt/Co/Ta MLs[33]. Nucleation fields above 100 mT were applied to obtain skyrmions with average diameter of 200 nm at remanence[33] on continuous MLs. For applications, it is desirable to be able to nucleate smaller skyrmions, and with the lowest magnetic field possible. Here we show that the specific topology of the nanodomes can help to break up the large magnetic domains and decrease the intensity of the magnetic field necessary to nucleate skyrmions, forming ordered arrays of 100 nm skyrmions at remanence.

We applied different reversal fields ($H_R$) of increasing intensity on the saturated sample to nucleate the skyrmions. Figures 3(a-c) show MFM images from the film on nanodomes with different out-of-plane applied fields. When a field $H_R$ = 50 mT is applied, it is possible to observe a dense array of circular individual skyrmions over the sample, with diameters of 100 nm in average. For a field of $H_R$ = 70 mT the density of skyrmions decreases and at $H_R$ = 90 mT only few skyrmions remain in the sample, as shown in Fig. 3(c). The density of skyrmions decreases with increasing applied field, as shown in the Fig. 3(h). Figures 3(d-f) show the MFM images obtained at remanence, after removal of the $H_R$ applied field. The curved topography of the film on nanodomes, with valleys between nanodomes that act as pinning sites for the propagation and merger of skyrmions when the field is removed, helps to stabilize dense arrays of skyrmions at remanence. In the MFM image obtained at zero field Fig. 3(d), after application of a 50 mT $H_R$, a dense array of hexagonally ordered skyrmions remain stable on the nanodomes. Some short stripe domains composed by chains of 100 nm circular domains are also observed in this image. Fig. 3(e) shows the MFM image taken at remanence after application of a 70 mT $H_R$, where the number of isolated skyrmions decreases and more worm-like domains can be observed. Finally, at remanence after application of a 90 mT $H_R$, most of the individual skyrmions merge into large worm domains, in a similar configuration as the one observed in the demagnetized state of Fig. 2(c). As shown in previous works[32, 33], when a large density of skyrmions are nucleated by appropriate field reversal protocols in the continuous MLs, the skyrmions are stabilized by mutual repulsion once a large skyrmion density is present in the sample[33], what prevents the merging of skyrmions into large domains after reducing the field to zero. Figure 3(g) shows a magnified view of a periodic lattice with hexagonally ordered skyrmions with an average diameter of 100 nm and separated by 200 nm. This indicates that for the ML on nanodomes, a skyrmion lattice pattern is formed to minimize the dipolar field energy and the additional effective energy contribution induced by geometrically broken symmetry, equivalent to maintain neighboring bits with anti-parallel magnetizations[34,35] on a recording media. Thus, the application of a magnetic field on an array of nanodomes would constitute an efficient mechanism to control the information bit density in data storage systems.

We have taken MFM images from the reference multilayer under similar conditions, and the results are shown in the figure S1 of the supplementary material. The MFM images from the film under an applied field of 50 mT shows a mixture of skyrmions (with average diameter of 200 nm) and short stripe domains. In the remanent state the stripe domains grow larger and only a few skyrmions remain stable. When a field of 70 mT is applied, the average skyrmion diameters decrease to 150 nm, and again they grow into large stripe domains at remanence. In a previous work[33] we only obtained stable skyrmions at remanence in films with thicker Co layers ($t_{Co}$= 1.7 nm), as consequence of a change in the effective anisotropy of the MLs with increasing Co thickness[33, 11].

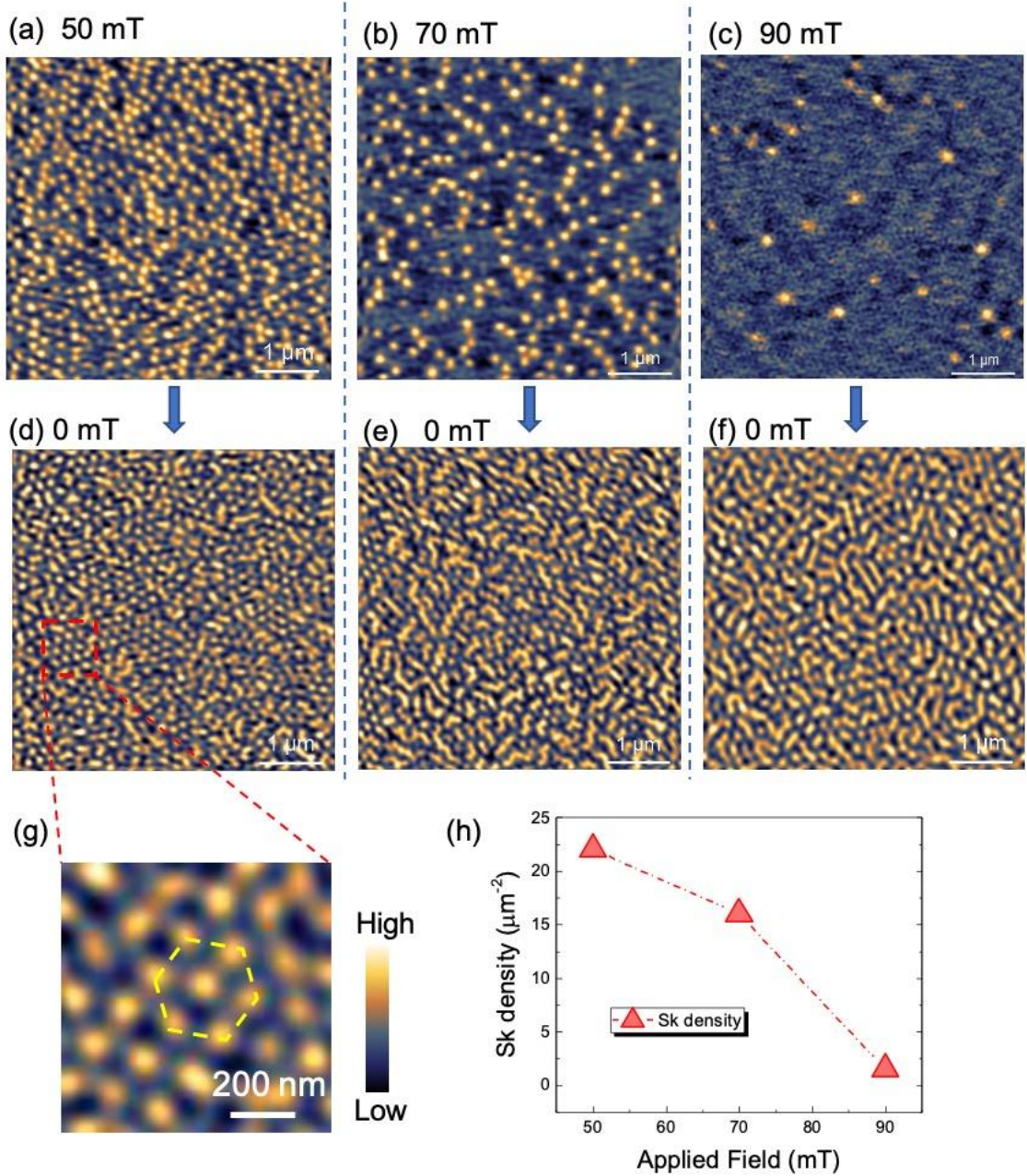

Fig. 3. MFM images of nanodomes under an out-of-plane applied field of (a) 50 mT, (b) 70 mT and (c) 90 mT. (d) to (f) MFM images of the nanodomes in the remanent states after the field was set to zero. (g) Magnified view of the skyrmion lattice marked in (d), with hexagonally ordered skyrmions. (h) Skyrmion density as a function of applied field for the MFM images from (a) to (c).

**Micromagnetic simulations of arrays of nanodomes.** We performed micromagnetic simulations with the Mumax3 program to understand how skyrmions are nucleated and stabilized in the multilayers. Fig. 4 shows the image of an initial magnetization state that was set to a Néel skyrmion and subsequently let to relax to equilibrium conditions, for a reference ML (Fig.4.a) and for the nanodome system (Fig.4.b). It can be clearly seen that in the reference ML the domain configuration evolves to a labyrinthine configuration, and in the nanodomes the curved topography prevents the skyrmion to grow larger than the dome diameter. Figure S2 in supporting information shows simulation images of the demagnetized state of the reference sample and from the nanodomes, where a narrower labyrinth domain configuration is observed in the nanodomes as compared to the reference multilayer, in complete agreement with the results observed on the MFM images of Fig.2.

To reproduce accurately the magnetic properties observed for the multilayers we simulated several hysteresis curves with the main magnetic parameters obtained from the experiment, until the simulated curve matched almost perfectly the experimental one, as shown in Fig. 4(c) for the film on nanodomes, and in Fig. S3 for the reference multilayer. The saturation magnetization value ($M_S = 0.85 \times 10^6$ A/m) used in the simulation was obtained from the experimental magnetization measurements, while the uniaxial anisotropy and the DMI constant were varied to determine the values that best matches the experimental results, that in the case of the film on nanodomes was $D = 1.3 \times 10^{-3}$ J m$^{-2}$ and $K_u = 0.54 \times 10^6$ J/m$^3$. The exchange constant was chosen to be $A = 1.0 \times 10^{-11}$ J/m and temperature was set to 300 K for 5 ns before the system was relaxed.

Figure S3 shows the simulated hysteresis curve and the simulated images from the reference multilayer under different out-of-plane applied fields and at remanence. While skyrmions are nucleated when the field is applied, at remanence they merge into large domains. The simulated domain configuration of the nanodomes, obtained after applying a magnetic field with intensity similar to the one used when obtaining the MFM images in Fig.3(a-c), is shown in Figures 4(d-f). It is possible to observe that the curved topography and the valleys between successive nanodomes are effective as pinning sites for domain wall propagation[19,20,34], keeping the domain size to the nanodomes' diameter. This behavior is explained by the effect of local periodicity produced by the curvature of the nanodomes. The application of an increasingly more intense magnetic field allows the sizes of the worm domains to decrease enough so the local effect of the curvature becomes important, allowing their transformation to a metastable skyrmion state. It also shows that the skyrmions' density decreases with increasing intensity of the magnetic field, and since the isolated skyrmions lie on the hexagonal lattice of nanodomes, their configuration is similar to the ones observed in the MFM images of Fig. 3(a-c), with isolated skyrmions separated by approximately 200 nm.

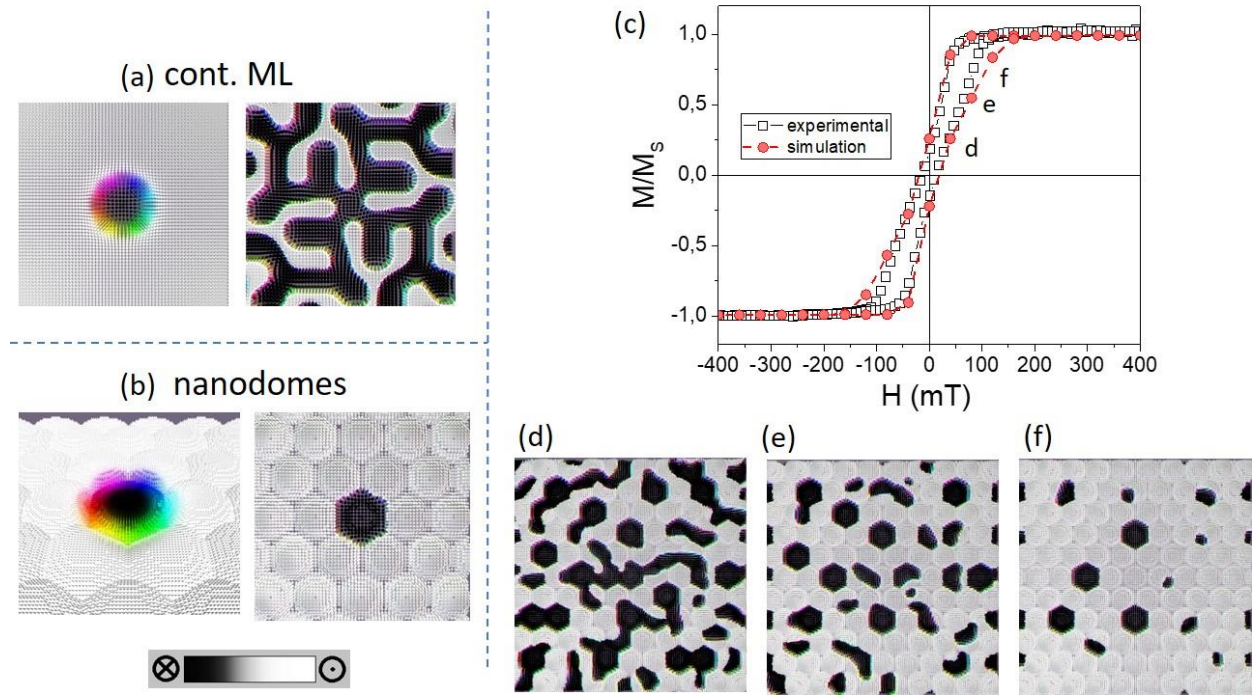

Fig. 4. Simulation of the initial magnetization state that was set to a Néel skyrmion, and subsequently let to relax to equilibrium conditions for a reference ML (a) and for the film on nanodomes (b). (c) Simulated hysteresis curve (red circles) and the experimental out-of-plane magnetization curve (black squares) for the film on nanodomes. (d) to (f) simulated domain configuration of the nanodomes obtained at the points d, e and f of the hysteresis curve in (c), for fields of 40, 80 and 120 mT respectively.

Additionally, we have simulated the application of a localized field on the individual structure of each nanodome, and we have been able to stabilize with high precision individual skyrmions on the saturated sample, as shown in Fig. 5a. Recent studies have shown that localized fields can be created by means of a magnetic tip in the proximity of a nanodot, which allows the creation and annihilation of isolated skyrmions[36-38]. In this way, our results suggest a mechanism for nucleate/annihilate a skyrmion, precisely controlling its location in an extended system of nanodomes. For example, starting with a uniformly magnetized sample, the application of a localized magnetic field in the opposite direction to the magnetization of the sample (which can be created by a magnetic tip) will nucleate a skyrmion on the nanodome. Subsequently, the magnetizing tip can continue to move over the sample and record new skyrmions. Fig. 5b shows a schematic representation of the recording mechanism. It is worth noting that the skyrmion core size can be controlled through the application of a magnetic field[39] so that the skyrmion core size increases (decreases) when the magnetic field is applied parallel (antiparallel) to the direction of its core. In this way, a suitable magnetic field allows for the skyrmion core to decrease enough to reconstruct the uniform magnetic state over the dome, constituting an efficient annihilation mechanism.

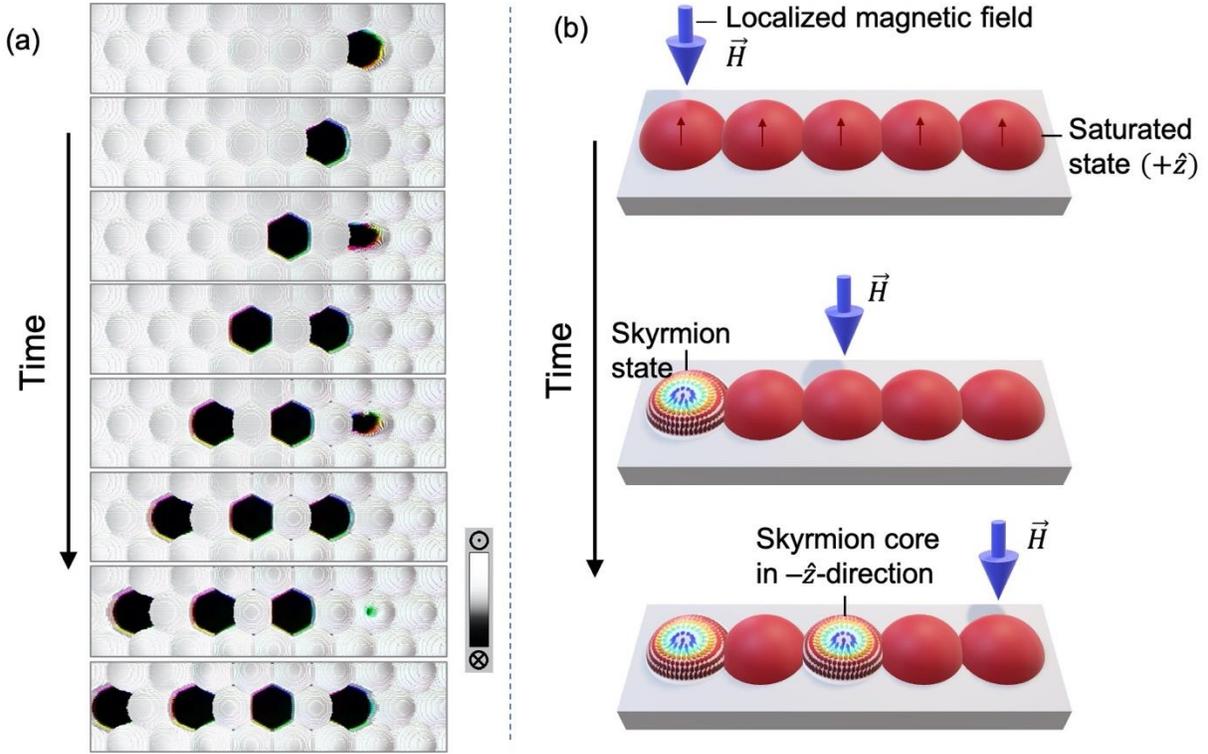

Fig. 5. Nucleation mechanism of isolated skyrmions on the surface of nanodomes. (a) A localized magnetic field is applied to the structure of a nanodome. Once the skyrmion is nucleated, the magnetic field changes its position to generate a new skyrmion on intercalated nanodomes. (b) Schematic representation of the skyrmion nucleation process presented in (a). The arrow represents a localized magnetic field, which can be created by a magnetic tip that moves through the sample.

**CONCLUSIONS**

In conclusion, we have shown for the first time the formation and stabilization of magnetic skyrmions on the confined geometry of hexagonal nanodomes fabricated in the barrier layer of alumina membranes. From the hysteresis loops we extracted the saturation magnetization $M_s$, anisotropy field $H_k$, and observed that the perpendicular magnetic anisotropy $K_{eff}$ decreased as the nanodome diameter increased. MFM images taken at the demagnetized state revealed that the topography of the nanodomes strongly affects the domain configuration of the film, and allows for the existence of smaller labyrinth domains, formed by thinnest stripes, through the entire sample. Large arrays of skyrmions are nucleated by the application of out-of-plane fields as small as 50 mT. The skyrmions stabilize at remanence in hexagonal ordered arrays, with diameters of 100 nm and separated by 200 nm. Agreement between the simulations and the experimental results was obtained, which indicates that the magnetic parameters obtained from magnetization measurements were appropriate. Self-assembled nanodomes have several advantages over arrays produced by lithographic based methods, including low-cost and extremely simple processing. The nanodomes can be tuned to diameters as small as 10 nm, which opens the possibility for further investigation of self-assembled substrates to be used in designing ultra-high density skyrmionic materials. Our results also suggest a mechanism for nucleate/annihilate skyrmions on nanodomes using a magnetizing probe, precisely controlling their location.

## ASSOCIATED CONTENT

**Supporting Information**

MFM images of the reference multilayers; simulations of the demagnetized state of reference ML and nanodomes; simulation results of reference ML with different applied fields and at remanence.

## AUTHOR INFORMATION

**Corresponding Author**

*E-mail: juliano.denardin@usach.cl

**ACKNOWLEDGMENT**

This study was financed in part by the Coordenação de Aperfeiçoamento de Pessoal de Nível Superior - Brasil (CAPES) - Finance Code 001. L.S.D. acknowledges support from CNPq grant 302950/2017-6. The authors would like to thank the financial support from Chilean Agencies Basal CEDENNA AFB180001, projects USA1899-2-3-2, 042031CD/Posdoc and CIP2018006 by Univ. Central de Chile, ANID/CONICYT/Fondecyt 1200782 and 1201491 and ANID-PFCHA/Postdoctorado Becas Chile 74200122.

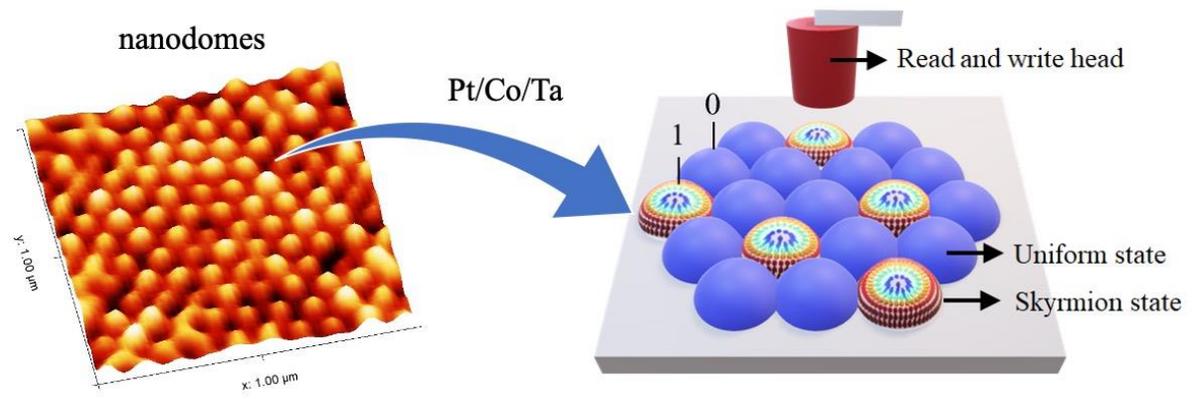

Table of Contents (TOC) graphic

**Supporting Information**

# Stabilization of Magnetic Skyrmions on Arrays of Self-Assembled Hexagonal Nanodomes for Magnetic Recording Applications

Felipe Tejo,[1,4,5] Denilson Toneto,[2] Simón Oyarzún,[1,4] José Hermosilla,[1] Caroline S. Danna,[1] Juan L. Palma,[3,4] Ricardo B. da Silva [2], Lucio S. Dorneles,[2] and Juliano C. Denardin[1,4,*]

[1]*Departamento de Física, Universidad de Santiago de Chile, Santiago, Chile.*

[2]*Departamento de Física, Universidade Federal de Santa Maria, UFSM, Santa Maria, RS, Brazil.*

[3]*Escuela de Ingeniería, Universidad Central de Chile, Santiago, Chile*

[4] *CEDENNA, Universidad de Santiago de Chile, Santiago, Chile.*

[5] *Instituto de Ciencia de Materiales de Madrid, CSIC, Cantoblanco, 28049 Madrid, Spain.*

Correspondence and requests for materials should be addressed to JCD (juliano.denardin@usach.cl)

Figure S1 shows MFM images from the film under an applied field of 50 mT, 70 mT and at remanence after the field is removed. Fig. S1(a) shows a mixture of skyrmions (with average diameter of 200 nm) and short stripe domains. In the remanent state the stripe domains grow larger and only a few skyrmions remain stable. When a field of 70 mT is applied, the average skyrmion diameter decreases to 150 nm, and again they grow into large stripe domains at remanence. At an applied field of 90 mT the film was saturated, and no magnetic contrast was observed in the MFM, the remanent state after saturation is shown in previous results[1] and is like the demagnetized state (Fig. 2.a).

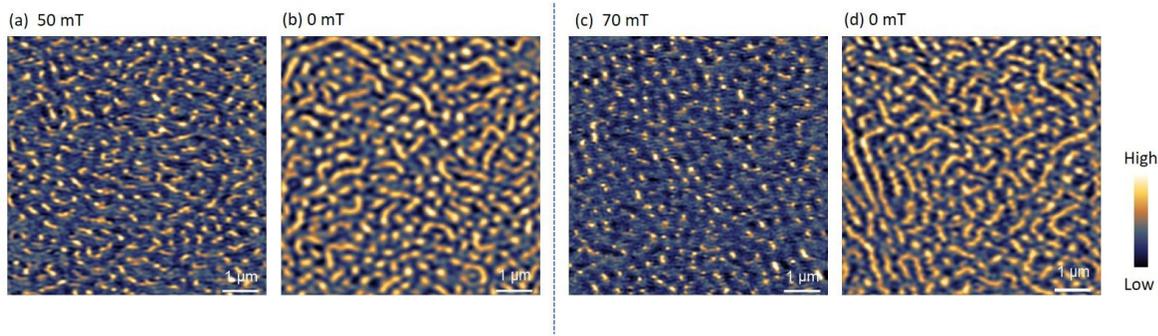

Figure S1. MFM images from the reference multilayer under an out-of-plane applied field of (a) 50 mT and (c) 70 mT. (b) and (d) MFM images of the reference multilayer in the remanent states after the field was set to zero.

The images from the simulated demagnetized state from the reference sample and from the film on nanodomes are shown in Fig. S2, where a narrower labyrinth domain configuration is observed in the film on nanodomes as compared to the reference multilayer, in complete agreement with the results observed on the MFM images of Fig.2. The simulation size was $2 \times 2$ μm$^2$, and the magnetic parameters where adjusted to obtain a hysteresis curve similar to the experimental one, as shown in Fig. S3 (a), with magnetic parameters of $M_S = 0.85 \times 10^6$ A/m, $K_u = 0.8 \times 10^6$ J/m$^3$ and $D = 0.3 \times 10^{-3}$ J m$^{-2}$. The exchange constant was chosen to be $A = 1.0 \times 10^{-11}$ J/m and temperature was set to 300 K for 5 ns before the system was relaxed. Fig. S3(b) shows the simulated images after different out-of-plane fields were applied and at remanence after the field was set to zero.

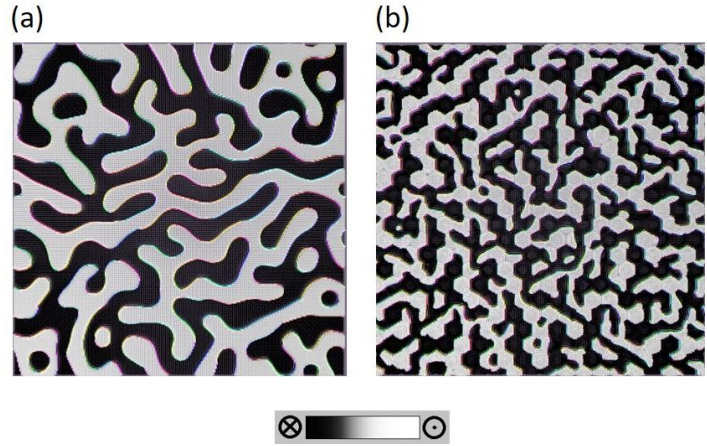

Figure S2. Simulation images of the demagnetized state of the (a) reference sample and (b) from the nanodomes.

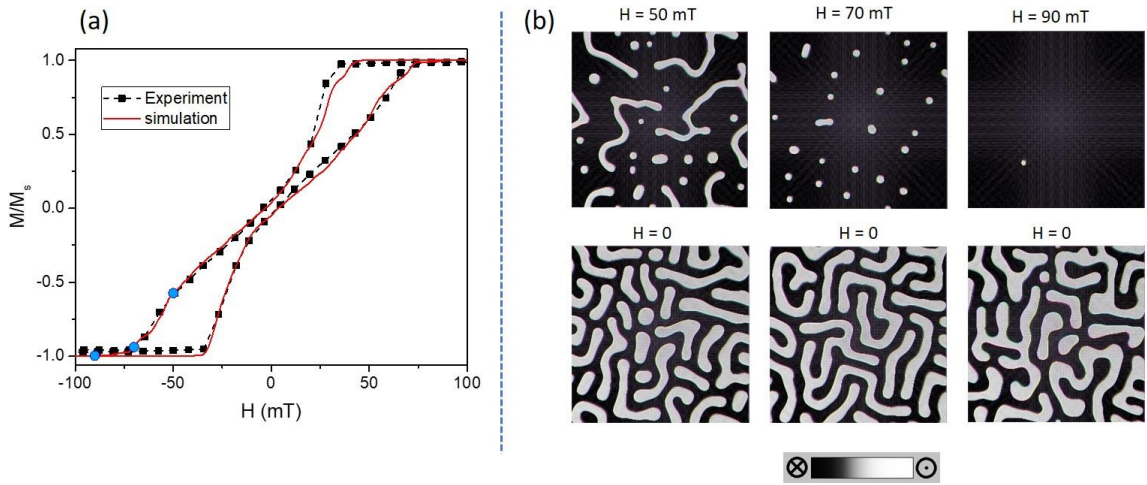

Figure S3. (a) Simulated hysteresis curve and (b) simulated images for the reference multilayer under different out-of-plane applied fields and at remanence.